\documentclass[a4paper,fleqn]{cas-sc}

\usepackage[authoryear,longnamesfirst]{natbib}

 \newdefinition{rmk}{Remark}
 \newproof{proof}{Proof}
 \usepackage{color,soul}

\begin{document}
\let\WriteBookmarks\relax
\def\floatpagepagefraction{1}
\def\textpagefraction{.001}

\shorttitle{A tunable population timer in multicellular consortia}
\shortauthors{Toscano-Ochoa and Garcia-Ojalvo}

\title[mode = title]{A tunable population timer in multicellular consortia}                      
%
%

\author{Carlos Toscano-Ochoa}[orcid=0000-0002-1239-0021]
\ead{biotoscano@gmail.com}

\author{Jordi Garcia-Ojalvo}[orcid=0000-0002-3716-7520]
\ead{jordi.g.ojalvo@upf.edu}

\address{Department of Experimental and Health Sciences, Universitat Pompeu Fabra,
Barcelona Biomedical Research Park, Dr. Aiguader 88, 08003 Barcelona, Spain}

%

\begin{abstract}
Processing time-dependent information requires cells to quantify the duration of past regulatory events and program the time span of future signals.
At the single-cell level, timer mechanisms can be implemented with genetic circuits: sets of genes connected to achieve specific tasks.
However, such systems are difficult to implement in single cells due to saturation in molecular components and stochasticity in the limited intracellular space.
Multicellular implementations, on the other hand, outsource some of the components of information-processing circuits to the extracellular space, and thereby might escape those constraints. 
Here we develop a theoretical framework, based on a trilinear coordinate representation, to study the collective behavior of a cellular population composed of three cell types under stationary conditions.
This framework reveals that distributing different processes (in our case the production, detection and degradation of a time-encoding signal) across distinct cell types enables the robust implementation of a multicellular timer. Our analysis also shows the circuit to be easily tunable by varying the cellular composition of the consortium.
\end{abstract}

%
%
\begin{keywords}
biological time encoding \sep chemical wiring problem \sep distributed computing \sep synthetic cell consortia \sep polybacterial communities \sep quorum sensing
\end{keywords}

\maketitle

\noindent

\section{Introduction}
One of the defining features of cells is their ability to process information, by way of sensing internal and external signals and generating a response.
This paradigm, while serving as an inspiration for the development of synthetic biology \citep{Andrianantoandro2006},  has also guided our search for gene and protein circuits responsible for all kinds of cellular functions \citep{Alon:2019aa}.
Frequently, studies of cellular circuits have focused on situations in which the input signal is constant throughout time.
However, cells usually live in time-dependent environments \citep{Tagkopoulos:2008aa}, and thus they face the need to process information that varies in time in a complex manner \citep{Gabalda-Sagarra:2018aa}.
An important instance of temporal information processing by cells is the quantification of the time interval that a given input signal has been acting upon a cell.
Complementarily, cells sending signals might need to program the amount of time that the signal is active.
In technological settings, both these functions are performed by timers.
It is thus necessary to establish how timers are implemented in cells.

Cell-intrinsic molecular timers have been identified in recent years, having been reported to regulate a wide variety of cellular processes, including apoptosis \citep{Hou:2004aa,Malladi:2009aa,Gerecht:2020aa,Fullstone:2020aa}, cellular proliferation \citep{Yeh:2007aa}, cell-fate specification \citep{Durand:2000aa,Raff:2007aa}, and infection response \citep{Eckert:2005aa}.
These timers usually depend on intricate molecular processes \citep{Lu:2007aa} that are difficult to tune and are sensitive to noise.
From a systems-level perspective, feedback-based timer circuits have been proposed in synthetic biology applications \citep{Ellis:2009aa,Pinto:2018aa}, and have also been found to operate naturally through pulses \citep{Sen:2011aa,Levine:2012aa} and oscillations \citep{Cai:2014aa}, and even to be induced by noise \citep{Turcotte:2008ab}.
These intracellular timer circuits usually depend on an accumulating signal surpassing a threshold \citep{Levine:2014aa}, and are thus limited by the amount of signal molecules that cells can produce and store in their interior without changing their basal metabolic state.
This limitation disappears if the factor controlling the timer is exported outside the cell and stored in the extracellular medium.
Cells in such multicellular timers would operate collectively, and consequently their function would be less affected by noise \citep{Enright:1980aa}.
In this paper we introduce a minimal implementation of a multicellular timer, using a novel theoretical framework to describe the collective dynamics of cellular populations, using microbial consortia as a specific example.

Microbial consortia are mixed populations of microorganisms, usually bacteria, that exhibit division of labor among different species or strains.
Extensive efforts have been devoted in recent years to engineer genetic circuits in synthetic consortia \citep{BRENNER2008,pai2009engineering}, pushing the development of distributed computation systems \citep{tamsir2011robust,regot2011distributed,sole2013expanding}.
Genetic circuits consist of several genes whose products interact to implement specific functions.
Under this approach, these circuits are split into different components (individual logic gates or small networks motifs), and implemented in distinct cell subpopulations. These subpopulations are then connected between them by using extracellular signals to transfer information.
This distributed approach reduces the amount of new genetic material that has to be introduced into a single cell type, and removes crosstalk between different circuit components by placing them in different cells.

Cell-cell communication in this context is usually based on the quorum sensing systems of gram-negative bacteria \citep{papenfort2016quorum}.
These systems are commonly based on an enzyme that produces AHL (acyl-homoserine lactone), a family of chemical compounds with the ability to diffuse through cellular membranes, bind to a transcription factor, and thereby regulate gene expression.
In the distributed computing scenario, the sender and receiver circuit modules that produce and sense AHL, respectively, are placed in different bacterial strains \citep{regot2011distributed,tamsir2011robust}.
Even though this sender-receiver framework allows to implement systems with the ability to self-regulate and process information collectively, they lack a component that actively degrades the chemical signal in the medium.
This implies that when the sender strain is turned off, the chemical substance that transfers the information may disappear very slowly if it is relatively stable.
This limits the ability of the circuit to process time-dependent information.
Therefore, the consortium should include a strain that degrades the chemical signal \citep{danino2010synchronized,prindle2012sensing,Silva:2017aa}.

Our cell consortium constitutes a minimal information transmission system that we call a chemical wire.
The system is composed of three different cell types (Fig.~\ref{fig:1}A): the first one produces a chemical compound that mediates the communication (emitter, E, green cells in the figure), the second detects the signal (receptor, R, blue cells), and the third actively degrades it (sink, S, red cells). 
In what follows, we establish a theoretical framework to describe this wire architecture, and provide mathematical definitions of what we call the wire space.
This framework is built with the aim to provide both a practical tool for bioengineers and a systems-level description of multicellular information transmission and processing.
Using this approach, we show that cell consortia can work as a timers, by remembering how long the sender strain has been in the ON state.
In turn, the time during which the receiver strain is active can be programmed by tuning the duration of the input pulse.
We consider for simplicity that the population is not structured in space and is well mixed (corresponding to a liquid culture), and that it operates in the stationary phase (i.e. the cells do not proliferate), although these conditions are not strictly required, as discussed in Sec.~\ref{sec:disc} below.
Additionally, stationarity is not a limiting assumption from the biochemical point of view, since it has been shown that bacteria such as \textit{Escherichia coli} are able to respond to external stimuli even in stationary phase \citep{Gefen2014}.
This establishes a less restrictive environment for the implementation of consortia circuits connected through chemical wires, since the relative growth rates of the different cell types involved will not play a relevant role in the operation of the system.
An earlier proposal of a multicellular timer involved a growing population of cells, but as a result it required the use of a single cell type, and the output was dominated by noise \citep{Koseska:2009ab}.
In what follows, we present a theoretical framework that allows us to analyze the behavior of our three-strain multicellular system quantitatively.

\section{Results}
\subsection{Wire definition and model for cultures of non-growing cells}

\begin{figure}[t]
\centering
\includegraphics[width=0.9\textwidth]{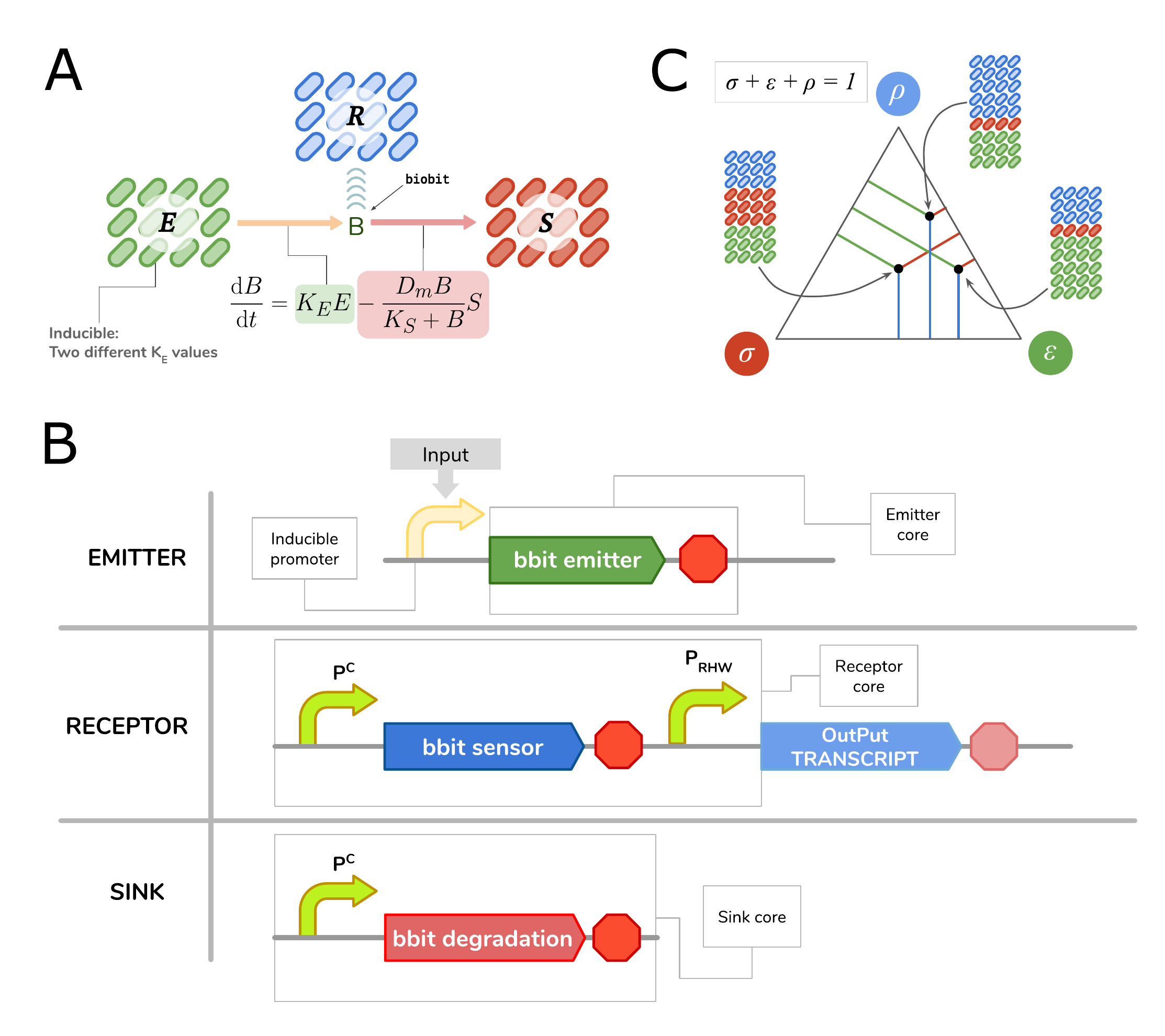}
\caption{\textbf{A minimal biochemical wire.} (A) Wire structure. The emitter strain E emits bbits to the medium at a constant rate, proportional to the strain density; the receptor strain R senses the bbit signal intensity; and the sink strain S degrades bbits with a Michaelis-Menten dynamics proportional to the strain density. \textit{B} stands for the bbit signal intensity, which will be bbit concentration in most cases. 
(B) Genetic constructs that implement the different elements of the wire. The emitter module (top) is the minimum set of genes that emit bbits when the corresponding promoter is induced.
The receptor module (middle) is the minimum set of genes (constitutively expressed from the $\mathrm{P^C}$ promoter) that sense the bbit and induce gene expression from the promoter $\mathrm{P_{R}}$ promoter, when the bbit concentration in the medium is above some threshold.
The sink module (bottom) is the minimum set of genes (constitutively expressed from the $\mathrm{P^C}$ promoter) that degrade bbit in the medium.
In the figure, the red stop signs stand for transcription terminators.
(C) Trilinear coordinate system.
Each point within the triangle represents a unique combination of emitter, receptor and sink.
\label{fig:1}}
\end{figure}

Our chemical wire is implemented by three genetic constructs (Fig.~\ref{fig:1}B), located in the three cell types defined above: an emitter, a receiver, and a sink.
These genetic constructs are chimeric genes composed by parts from different genes in order to accomplish the desired tasks.
We call the communicating signal bbit, which stands for biological bit, or biobit.
This candidate molecule must be stable enough to consider it has a negligible abiotic degradation rate.
In our case the bbit will be the quorum sensing molecule AHL.
A bbit can take any value within the [0,1] interval depending on the input signal, and as we will see it can also store input duration.
The emitter construct contains the minimum set of genes required to synthesize bbit at a constant rate.
The receiver construct contains comprises the minimum set of genes required to induce gene expression from a promoter when bbit is present in the medium at a concentration above/below a specific threshold (depending on whether AHL activates or represses its target gene, respectively).
Finally, the sink component consists of the minimum set of genes, constitutively expressed, required to degrade bbit.
Our first goal is to analyze how these strains communicate with each other, and how the relative frequencies between them affect this communication.

We assume that the production of bbit, with concentration $B$, is linearly proportional to the density of the emitter strain.
In contrast, the sink circuit is be supposed to degrade bbit following a Michaelis-Menten dynamics, proportional to the sink strain density $S$ (Fig.~\ref{fig:1}A):
\begin{equation}
\frac{dB}{dt}=K_E E - \frac{D_m B}{K_S + B}S
\label{eq:Bdyn}
\end{equation}
The Michaelis-Menten term in the equation above represents the limited ability of the sink strain to degrade bbit, which arises when the bbit concentration $B$ becomes large enough to saturate the sink cells.
This saturation is quantified by the maximum degradation rate $D_m$, and starts to be manifested when the biobit concentration $B$ surpasses the Michaelis constant $K_S$. This last constant depends on the affinity of the bbit to its degrading enzyme, which is in turn expressed by the sink cells.
Also, in this term the density $S$ of sink cells plays the role of the total ``enzime'' concentration of the Michaelis-Menten kinetics, and thus appears multiplying the degradation rate $D_m$.

In this equation $E$ and $S$ represent the concentration of emitter and sink cells, respectively, while below we will also introduce the concentration $R$ of receptor cells.
These concentrations can be given in cells/mL or directly in optical density units (which could be more convenient when designing an experimental implementation of the system).
For simplicity, our wire will be conceived in stationary phase, a condition that still allows cells to respond to external stimuli \citep{Gefen2014}. This means that cells are alive but are not dividing nor growing, and there is no dilution of bbit.
We also assume that intrinsic bbit degradation is negligible in the time scales considered here.

It is convenient to define the relative frequency of each strain in the population: $\varepsilon=E/P$, $\sigma=S/P$, and $\rho=R/ P$, where $P$ is the total population density (which is constant provided the population is in the stationary phase).
By definition, $\varepsilon + \sigma + \rho =1$.
We now establish a bijection between any possible bacterial consortium and a point inside an equilateral triangle (Fig.~\ref{fig:1}C).
In this representation, each vertex in the triangle corresponds to one of the three strains, and each of the relative frequencies $\varepsilon$, $\sigma$ and $\rho$ is given by the distance between the point representing the consortium and the side of the triangle opposite the corresponding vertex\footnote{The triangle must have equal sides to obey the constraint that the sum of the distances from any point lying within it to its three sides is constant (equal to 1 in our case, as we have seen above).}.
This is called a trilinear coordinate representation, and it will be our graphical framework to explore the behavior of our bacterial consortium.
The intuition behind this geometrical representation is simple: the closer the system is to a vertex, the more represented the corresponding strain will be in the consortium.
Since the population is in the stationary phase, none of the bacterial strains change in density, and the point describing the system will be fixed inside the triangle.
We define the set of all possible consortia as the wire space $\Delta$ (the interior of the triangle).
Any specific point within the wire space will correspond to a unique population, whose behavior we will explore in the following sections. 

\subsection{Characterizing the bbit build-up}

\begin{figure}[htb]
\includegraphics[width=\textwidth]{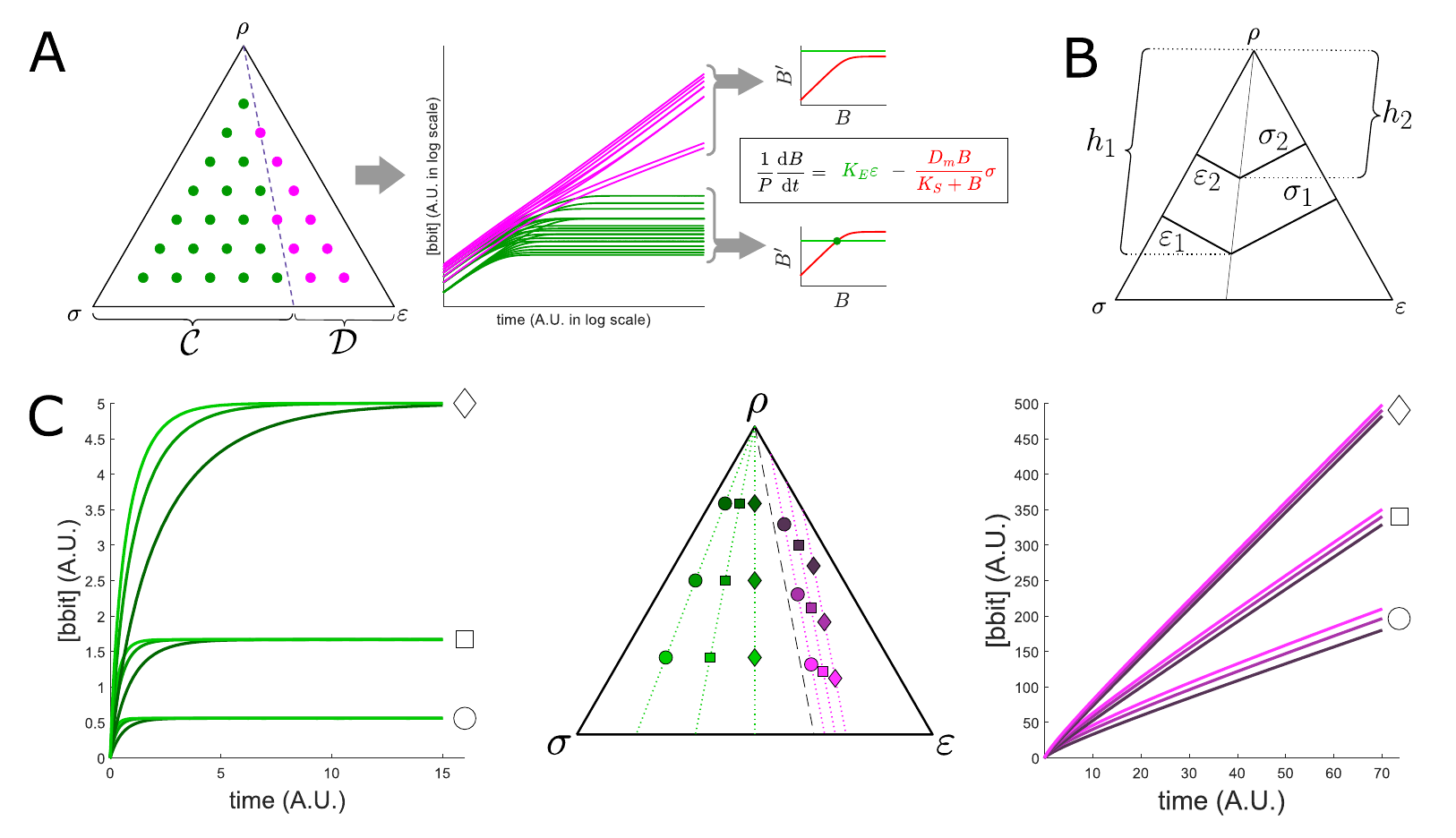}
\caption{\textbf{Dynamical dissection of the wire space}. (A) All the points lying in the convergence region (green dots in wire space, left) are associated with convergent bbit time-traces (center, green lines).
This is due to the intersection between the bbit production term and the bbit degradation term of Eq.~(\ref{eq:Bdyn}) (right, bottom inset).
Meanwhile, all points lying in the divergence region (magenta dots in wire space, left), are associated with quasi-linear bbit growth (center, magenta lines).
This is due to the lack of intersection between the bbit production term and the bbit degradation term of Eq.~(\ref{eq:Bdyn}) (right, top inset).
(B) Graphical support for the $B_{eq}$ Isolines Lemma: $(\sigma,\varepsilon)$ values for two points along an isoline define two similar triangles.
(C) Isolines: when cultures are placed on the same line radiating from the top vertex to the bottom side within the convergence region, they reach the same $B_{eq}$ value (green dots and lines).
In the divergence region, lines parallel to the divergence horizon have the same asymptotic growth rate of $B(t)$.
Parameter values for all simulations shown here: $K_E = 0.5$, $K_S = 5$, $D_m = 1$, and $P = 5$.
}
\label{fig:2}
\end{figure}

We can calculate the fixed points of Eq.~(\ref{eq:Bdyn}) to obtain the explicit expression of bbit concentration at equilibrium, which depends on the fractions of the emitter and sink cells, $\varepsilon$ and $\sigma$ respectively:
\begin{equation}
B_{eq}=\frac{K_E K_S \varepsilon}{D_m \sigma - K_E \varepsilon}
\label{eq:Beq}
\end{equation}
The pairs $(\varepsilon, \sigma)$ that make the denominator in $B_{eq}$ equal to zero divide the wire space in two regions (see dashed line inside the triangle in the left panel of Fig.~\ref{fig:2}A). 
When $D_m \sigma > K_E \varepsilon$ (region at the left of the line, covered by green dots), the bbit concentration reaches the equilibrium value $B_{eq}$ defined by Eq.~(\ref{eq:Beq}).
We call this the \textbf{convergence region} (represented by $\mathcal{C}$ below).
Solving the differential equation (\ref{eq:Bdyn}) in time within the convergence region shows that the bbit time traces converge to their predicted equilibrium concentration (green lines in the middle panel of Fig.~\ref{fig:2}A).
This corresponds to a situation in which the production of the bbit by the emitter strain is balanced by its degradation by the sink strain, as shown in the bottom inset in the right side of Fig.~\ref{fig:2}A.
The relative values of the production and degradation terms (green and red lines in the inset) at the two sides of the fixed point (production larger than degradation below the equilibrium, and degradation smaller than production above it) ensure that the equilibrium is stable.
In contrast, when $D_m \sigma < K_E \varepsilon$ (region at the right of the dashed line in Fig.~\ref{fig:2}A, covered by magenta dots), the production of bbit by the emitter strain is always larger than (and thus cannot be balanced by) the saturated degradation by the sink strain (top inset in the right side of Fig.~\ref{fig:2}A).
This leads asymptotically to a linear growth of $B$ in time (magenta lines in the middle panel of Fig.~\ref{fig:2}A; note that these lines are not linear from the beginning).
We call the region in which $D_m \sigma < K_E \varepsilon$ the \textbf{divergence region}\footnote{We note that in the divergence region, bbit production would eventually stop due to limited capacity of the medium, toxicity, lack of precursors, etc. We ignore these limitations in what follows.} $\mathcal{D}$.
These two regions are separated by what is called the \textbf{horizon of divergence} (represented by $\hat{\mathcal{H}}$).
Together, the three regions $\mathcal{C}$, $\mathcal{D}$, and $\hat{\mathcal{H}}$ comprise the wire space $\Delta$.
The mathematical characterization of these regions is summarized in Table~\ref{tab:1}. 

Within the convergence region, all points along any straight line radiating from the top $\rho$ vertex of the wire space triangle and hitting its $\overline{\sigma \varepsilon}$ bottom side, reach the same $B_{eq}$ value.
This can be shown by noting that the expression (\ref{eq:Beq}) of the bbit concentration at equilibrium can be rewritten as $B_{eq}=\frac{K_E K_S }{D_m \sigma/\varepsilon - K_E}$. Thus, $B_{eq}$ depends on the consortium composition only via the ratio $\sigma/\varepsilon$. In turn, geometric similarity shows in a straightforward manner that the ratio $\sigma/\varepsilon$ is constant for all points belonging to any straight line radiating from the $\rho$ vertex of the triangle.
To see this, consider two points along one of these radiating lines, characterized respectively by strain abundance fractions (relative frequencies) $(\sigma_1,\varepsilon_1)$ and $(\sigma_2,\varepsilon_2)$, as shown in Fig.~\ref{fig:2}B. Let $h_1$ and $h_2$ be distances between these two points and the $\rho$ vertex. The similarity between the two triangles delimited by the $\varepsilon$-$h$ sides in one case, and by the $\sigma$-$h$ sides in the other, leads to:
\begin{equation}
\frac{\varepsilon_1}{\varepsilon_2}=\frac{h_1}{h_2}=\frac{\sigma_1}{\sigma_2}
\quad\Longrightarrow\quad \frac{\sigma_1}{\varepsilon_1}=\frac{\sigma_2}{\varepsilon_2}
\quad\Longrightarrow\quad B_{eq, 1}=B_{eq, 2}
\label{eq:sim}
\end{equation}
The green lines and dots in the left and center panels of Fig.~\ref{fig:2}C depict some isolines and the behavior of $B(t)$ along them.

\begin{center}
\begin{table}[htb]
\centering

\begin{tabular}{ c c c c } 
\multirow{1}{2cm}{} &
 \multirow{1}{4cm}{Convergence Region} &
 \multirow{1}{4cm}{Horizon of Divergence} &
 \multirow{1}{4cm}{Divergence Region} \\
\hline
\multirow{2}{2cm}{Notation} &
 \multirow{2}{4cm}{$\mathcal{C}$} &
 \multirow{2}{4cm}{$\hat{\mathcal{H}}$} &
 \multirow{2}{4cm}{$\mathcal{D}$} \\
 & & & \\
\hline
 \multirow{3}{2cm}{Definition} &
 \multirow{3}{4cm}{${\displaystyle \frac{\varepsilon}{\sigma} < \frac{D_m}{K_E}}$} &
 \multirow{3}{4cm}{${\displaystyle\frac{\varepsilon}{\sigma} = \frac{D_m}{K_E}}$} &
 \multirow{3}{4cm}{${\displaystyle\frac{\varepsilon}{\sigma} > \frac{D_m}{K_E}}$} \\
& & & \\
& & & \\
\hline
 \multirow{3}{2cm}{Property} &
 \multirow{3}{4cm}{${\displaystyle B_{eq} = \frac{K_E K_S \varepsilon}{D_m \sigma - K_E \varepsilon}}$} &
 \multirow{3}{4cm}{${\displaystyle \left(\sigma, \varepsilon, \rho \right) = \left( k \hat{\sigma}, k \hat{\varepsilon}, 1-k \right)}$} &
 \multirow{3}{4cm}{${\displaystyle \hat{v} = P (D_m \sigma - K_E \varepsilon)}$} \\
& & & \\
& & & \\
\hline

\end{tabular}

\caption{Definition and properties of the convergence and divergence regions.
$\hat{\sigma}=\frac{K_E}{D_m+K_E}$ and $\hat{\varepsilon}=\frac{D_m}{D_m+K_E}$ are specific values of $\sigma$ and $\varepsilon$ that make the denominator of $B_{eq}$ equal to 0 with $\rho$ being 0, and $k$ is an arbitrary real number between 0 and 1.
\label{tab:1}}
\label{table:1}
\end{table}
\end{center}

Within the divergence region, each culture-point can be characterized by the asymptotic rate $\hat{v}$ at which $B(t)$ grows in time. Considering that $B(t)$ has grown well past $K_S$, the degradation term in Eq.~(\ref{eq:Bdyn}) becomes linear, leading to
\begin{equation}
    \hat{v} = P(K_E \varepsilon - D_m \sigma)
    \label{eq:vhat}
\end{equation}
We can ask which are the points in the divergence region that have the same $\hat{v}$ values.
Geometric considerations show that all points lying in a segment parallel to the horizon of divergence will have the same asymptotic growth speed given by Eq.~(\ref{eq:vhat}).
These \textit{growth-rate isolines} are shown in the right and center panels of Fig.~\ref{fig:2}C (purple lines and dots).

There are simple rules to gain intuition about both the convergence and divergence regions.
Within the convergence region, for instance, the closer an isoline is to the horizon of divergence, the greater its associated $B_{eq}$ value will be.
And within the divergence region, the closer a growth-rate isoline is to the $\varepsilon$ vertex, the greater its $\hat{v}$ value will be.
At the horizon of divergence, $\hat{v}=0$.
The trilinear coordinate representation introduced here provides us with an intuitive understanding of this multicellular information-processing system.
We now examine the response of this system to a modulation of the activity of the emitter strain, and argue that in a significant portion of the wire space this response can take the form of a timer.

\subsection{Wire ON/OFF states define digital and buffer regions}
\label{sec:4}

\begin{figure}[htb]
\includegraphics[width=\textwidth]{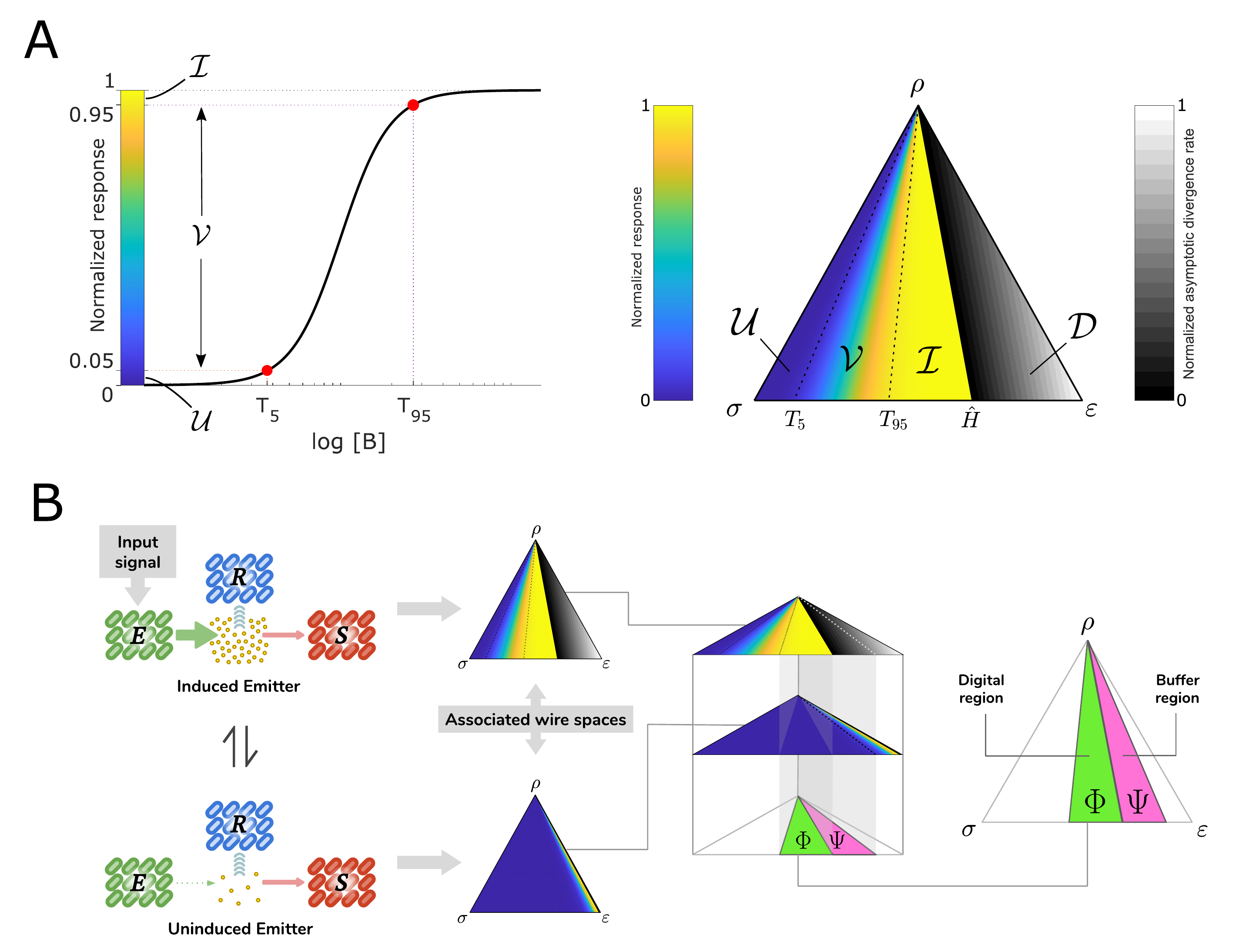}
\caption{\textbf{Response characteristics and emitter modulation}. (A) Transfer function associated to the receptor's response (left panel).
Two bbit concentration thresholds are fixed to define (right panel) when the receptor is uninduced ($B_{eq} \leq T_5$, region $\mathcal{U}$), when it is fully induced ($B_{eq} \geq T_{95}$, region $\mathcal{I}$), or when it has a variable induction level $B_{eq} \in (T_5, T_{95})$, region $\mathcal{V}$.
These three regions span the convergence domain, and are color coded according to the value of $B_{eq}$.
The rest of the wire space is covered by the divergence domain $\mathcal{D}$.
Within that region, color coding corresponds to the asymptotic accumulation rate of $B(t)$.
These regions structure the wire space in a way that informs us about the response of the wire given a set of parameters.
(B) When two states of the wire alternate (induced and uninduced emitter), two new regions can be defined.
The digital region $\Phi$ is the intersection of the induced region $\mathcal{I}$  (where the receptor is fully induced) when the emitter is ON, with the uninduced region $\mathcal{U}$ (where the receptor is uninduced) when the emitter is OFF.
The buffer region $\Psi$ is the intersection of the divergence region $\mathcal{D}$ (where bbit accumulates) when the emitter is induced, with the uninduced region $\mathcal{U}$ when the emitter is OFF.
Parameter values for all simulations shown here: $K_E^{\text{ON}} = 0.5$, $K_E^{\text{OFF}} = 0.01$, $K_S = 5$, $D_m = 1$, and $P = 5$. Hill function parameters: $K = 1$, with Hill coefficient $n = 3$.}
\label{fig:3}
\end{figure}

So far we have not considered the response of the receptor strain to the bbit present in the medium.
As discussed above (Fig.~\ref{fig:1}B), in that strain bbit induces the expression of an output protein, such as a fluorescent protein.
We assume that the transfer function relating the bbit concentration with its output follows an activating sigmoidal dynamics, described for instance by a Hill function:
\begin{equation}
F=\frac{B^n}{K^n+B^n}
\end{equation}
This function is plotted in the left plot of Fig.~\ref{fig:3}A, and entails assuming that the response of the output protein is fast in comparison the the other time scale of the dynamics.
We also normalize the response within the interval $(0,1)$.

We define two arbitrary thresholds for the receptor response, $T_5$ and $T_{95}$, corresponding to the bbit concentrations at which the receptor cells reach 5\% and 95\% of their maximum response, respectively.
These thresholds could be measured precisely in the experimental implementation of this system by quantifying its response. 
This allows us to ignore the heterogeneity existing not only in the response of individual receptor cells, but also in the bbit production rate of individual emitter cells.
Since all emitter cells contribute with the same communication molecule (the chosen bbit) to the medium, the bbit signal is averaged across the population, and thus this model is not affected by single-cell heterogeneity.

The receptor thresholds defined above, $T_5$ and $T_{95}$, allow us to split the convergence domain into three regions according to the induction level of the receptor (Fig.~\ref{fig:3}A, right):
(i) the uninduced region, named $\mathcal{U}$, in which $B_{eq} \leq T_5$ so that the receptor cells do not produce  any measurable output;
(ii) the induced region, named $\mathcal{I}$,  where $B_{eq} \geq T_{95}$ corresponding to a saturated maximal output readout;
and (iii) the variable region in between, named $\mathcal{V}$, where receptor cells reach a variable induction level with $B_{eq} \in (T_5, T_{95})$.
Within the divergence domain, receptor cells eventually reach the maximum level of induction upon constant activation of the emitter strain, but in this region the bbit concentration keeps accumulating at time goes on, in contrast with the induced region $\mathcal{I}$ in which bbit reaches an equilibrium value, as discussed above (this fact will be key for the timer operation to be discussed in the following Section).

So far we have characterized the wire when the emitter is constantly active.
But wires have two states by definition: ON and OFF, corresponding to the emitter being activated ($K_{E}^{\text{ON}} > 0$) and inactivated ($K_{E}^{\text{OFF}} \approx 0$), respectively.
These two wire states have different responses in the four regions defined above ($\mathcal{U}$, $\mathcal{V}$, $\mathcal{I}$, and $\mathcal{D}$) because the parameter $K_E$ changes between the two states, as shown in Fig.~\ref{fig:3}B).
If the difference between $K_{E}^{\text{ON}}$ and $K_{E}^{\text{OFF}}$ is large enough, interesting intersections between the two states can emerge:
\begin{itemize}
\item The \textit{digital region} $\Phi$ is the intersection between the $\mathcal{U}^{\text{OFF}}$ region (uninduced region $\mathcal{U}$ when the emitter is OFF), and the $\mathcal{I}^{\text{ON}}$ region (induced region $\mathcal{I}$ when the emitter is ON).
Within this region the consortium behaves in a fully digital manner, transducing directly into the response module/strain the input driving the emitter module/strain.
\item A more interesting region is the \textit{buffer region} $\Psi$, which corresponds to the intersection between the $\mathcal{U}^{\text{OFF}}$ region (uninduced region $\mathcal{U}$ when the emitter is OFF), and the $\mathcal{D}^{\text{ON}}$ region (divergence region $\mathcal{D}$ when the emitter is ON).
As we will see in what follows, within that region the response strain encodes the time during which the emitter strain has been ON.
Concurrently, the time during which the response is active can be programmed by fixing the duration of the ON state of the input.
In other words, within the $\Psi$ region the consortium can behave as a timer.
\end{itemize}

\subsection{The buffer region operates as a timer}
\begin{figure}[htb]
\includegraphics[width=\textwidth]{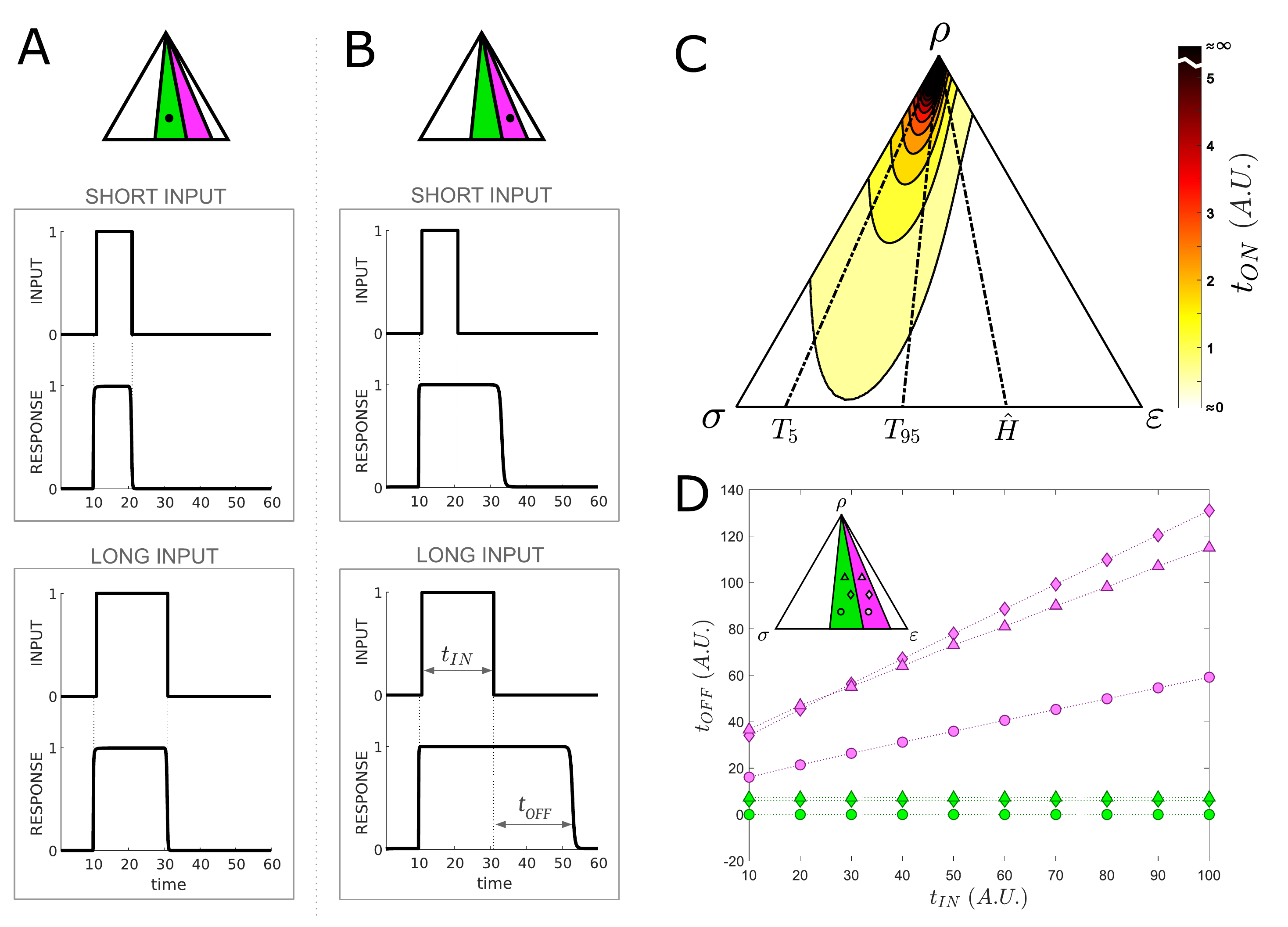}
\caption{\textbf{Digital response versus timer function.} (A) Response of the consortium to an input signal with two different durations when the culture is in the digital region.
(B) Corresponding behavior in the buffer region.
(C) Response time $t_{\text{ON}}$ across the wire space.
Darker colors correspond to a slower responses.
(D) Dependence of the turn-off time $t_{\text{OFF}}$ on the input duration ($t_{\text{IN}}$) for both the digital (green) and buffer (magenta) regions.
Three different culture points are shown in each region, represented by symbols of different shapes.
Parameters used for these simulation are those of Figure 3. Simulations were done with custom code in MATLAB using ODE45 function.}
\label{fig:4}
\end{figure}

The two regions described in the previous section have different and useful behaviors.
First we focus on the digital region $\Phi$.
All culture-points within this region will behave in a digital manner, meaning that when the emitter is ON, the receptor will be fully induced.
In turn, when the emitter changes its state to OFF, then the receptor changes its state to fully uninduced.
A key point of this digital region is that the time that it takes to transition from the uninduced state to the induced state and, most importantly, the time needed for the reporter to become uninduced again after the emitter is turned OFF is constant anywhere within this region for an arbitrary culture-point within it, and independent of how long the emitter is ON (Fig.~\ref{fig:4}A).
This is the crucial feature that allows the system to behave in a digital manner (which will not be fulfilled in the buffer region, and we will see below).

A potential limitation of the digital behavior of the system in region $\Phi$ is the time needed for the system to react when the emitter is turned on.
For the parameters chosen in Fig.~\ref{fig:4}A that time is negligible in comparison with the duration of the pulse, but it is worth asking how this turn-on time depends on the location of the consortium within the wire space.
To address this question, we define the time $t_{\text{ON}}$ that it takes for the receptor to reach 95\% of its corresponding $B_{eq}$.
This quantity cannot be calculated analytically, but it can be computed systematically by simulating the behavior of the system for culture-points equally distributed over the wire space.
The results are plotted in Fig.~\ref{fig:4}C, which shows that this turn-on time is low for a large region of the wire space, diverging to infinity only when the culture is close to the $\rho$ vertex (where the receptor strain dominates over the other two).
In the bottom region of the wire space, the response is the fastest where the emitter strain dominates (i.e. most of the consortium is formed by emitter cells).
Supplementary Video 1 shows how the response gets delayed, eventually preventing receptor cells to get fully induced. This supplementary video also emphasizes the behavior of the wire for cultures lying on top of the same isoline.

Within the buffer region $\Psi$, all culture-points also reach maximum receptor induction when the emitter is turned ON, in a time that is fixed (i.e. is independent of the duration of the preceding OFF state) and given as well by Fig.~\ref{fig:4}C as discussed above.
However, in striking contrast with the digital region, the time that the system takes to transition back from the induced to the uninduced state when the emitter strain is turned OFF is not fixed, but crucially depends on the duration of the previous ON state.
In other words, the longer the ON state of the emitter, the longer it will take for the receptor to transition from the induced to the uninduced state (Fig.~\ref{fig:4}B).
This prevents the use of the system as a digital encoder within this region, but enables a second, potentially more interesting, ability, namely measuring the time that the ON signal was present (or conversely, programming the time that the receptor is going to be induced, by tuning the time during with the emitter is ON). See Supplementary Video 2 to better appreciate the difference between buffer and digital regions.

To quantify this behavior, we define $t_{\text{IN}}$ as the duration of the input pulse, and introduce the magnitude $t_{\text{OFF}}$ to measure how long it takes for the wire receptor to transition from the induced to the uninduced state, starting when the input signal disappears, and ending when the receptor reaches again a response less or equal to 5\% of its maximum response value at the uninduced state (see double-headed arrows in Fig.~\ref{fig:4}B).
Figure~\ref{fig:4}D compares the behavior of this magnitude in the two regimes: whereas in the digital region $t_{\text{OFF}}$ is not affected when $t_{\text{IN}}$ changes (as we anticipated above), within the buffer region the response pulse is extended a time $t_{\text{OFF}}$ that increases linearly in response to increasing $t_{\text{IN}}$, thereby providing the population with a clean control of the response time duration.
This linear dependence can be understood from the fact that, in the buffer region, the bbit accumulates linearly with time. Specifically, the dynamics (\ref{eq:Bdyn}) of the bbit concentration while the emitter is ON can be approximated asymptotically (when $B\gg K_S$) by
\begin{equation}
    \frac{dB}{dt}\approx K_{E}^{\text{ON}} E - D_m S
    \;\Longrightarrow\; B(t)\approx P(K_{E}^{\text{ON}} \varepsilon - D_m \sigma)t, 
    \label{eq:Bdyn2}
\end{equation}
where we have assumed that the bbit concentration is approximately zero at the beginning of the input pulse.
According to the expression above, the bbit concentration at the end of the input pulse is $B_{\text{ON}}\approx P(K_{E}^{\text{ON}} \varepsilon - D_m \sigma)t_{\text{IN}}$.
After the input pulse has ended (but while $B$ is still large in comparison with $K_S$), the dynamics of $B$ is given by:
\begin{equation}
    \frac{dB}{dt}\approx - D_m S
    \;\Longrightarrow\; B(t)\approx B_{\text{ON}}-PD_m \sigma t=
    P(K_{E}^{\text{ON}} \varepsilon - D_m \sigma)t_{\text{IN}} -PD_m \sigma t
    \label{eq:Bdyn3}
\end{equation}
The time $t_{\text{OFF}}$ taken by the response to come back down after the input pulse has ended can be calculated as the time needed by $B(t)$ in Eq.~(\ref{eq:Bdyn3}) to reach a low enough value $B_{\text{OFF}}$ (which can still be large in comparison with $K_S$), so that expression (\ref{eq:Bdyn3}) still holds:
\begin{equation}
    B_{\text{OFF}} = P(K_{E}^{\text{ON}} \varepsilon - D_m \sigma)t_{\text{IN}} -PD_m \sigma t_{\text{OFF}}
    \;\Longrightarrow\;
    t_{\text{OFF}}= \frac{B_{\text{OFF}}}{PD_m \sigma}+
    \left(\frac{K_{E}^{\text{ON}} \varepsilon}{D_m\sigma}-1\right)t_{\text{IN}}
    \label{eq:Bdyn4}
\end{equation}
This corresponds to an approximately linear dependence of $t_{\text{OFF}}$ on $t_{\text{IN}}$ in the buffer region, as shown in Fig.~\ref{fig:4}D. Also, when the culture is in the buffer region and $t_\text{IN}$ is fixed, the slope of $t_\text{OFF}$ solely depends on the $\varepsilon/\sigma$ ratio, as shown in Eq.~(\ref{eq:Bdyn4}). This explains the different slopes in Fig.~\ref{fig:4}D (magenta lines).
The analysis above shows that this linear dependence stems from (i) the approximately linear accumulation of $B$ with time, and (ii) the linear dependence of $t_{\text{OFF}}$ on the amount of bbit existing in the media when the input pulse ends, $B_{\text{ON}}$ (Eq.~\ref{eq:Bdyn3}).

The fact that $t_{\text{OFF}}$ remains fixed but greater than zero when the culture is within the digital region associates a lag time to that region, whereas in the buffer region the linear dependence on $t_{\text{IN}}$ dominates.
Therefore, biological wires can store information about the time they have been in the ON state when the relative ratios of the different strains forming the consortium lie within the buffer region. This establishes the underlying principle of the collective timer behavior.

\subsection{Filter response to periodic inputs within the buffer region}
\begin{figure}[htb]
\includegraphics[width=\textwidth]{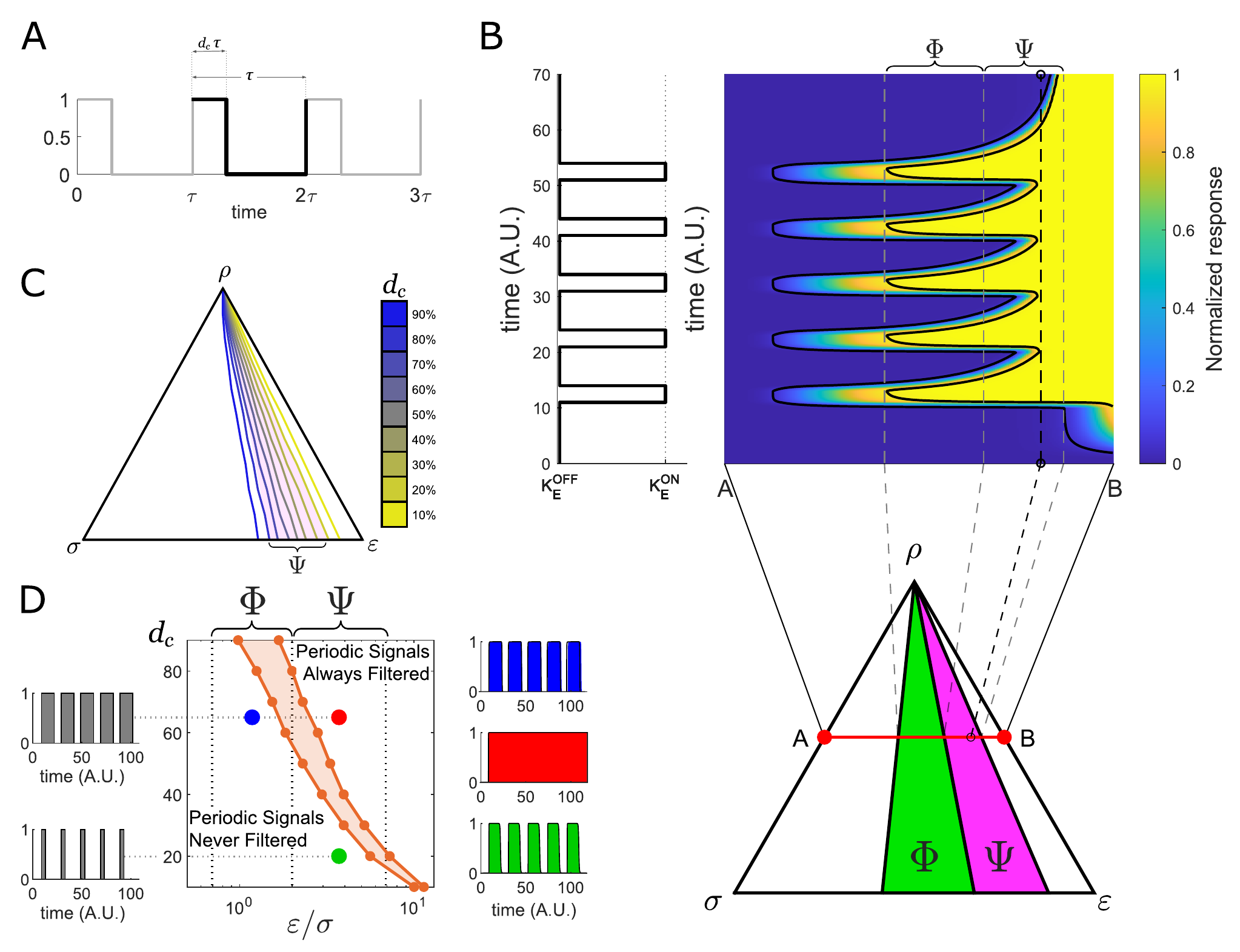}
\caption{\textbf{Filtering periodic pulses.} (A) Graphical definition of duty cycle ($d_c$): given a digital periodic signal acting as input signal, the duty cycle is the percentage of the period ($\tau$) that is in the ON state.
(B) Simulations of the receptor state for different culture locations within the wire space (color plot) for a fixed period of 10 time units and a 30\% duty cycle of the pulse train.
Solid black lines show where the $T_{5}$ and $T_{95}$ are crossed in the simulations.
The figure on the left represents the input signal.
The bottom figure shows the wire space with the digital (green) and buffer (magenta) regions.
The vertical black dashed line on the color plot shows the ratio $\varepsilon / \sigma$ at which the input signal is filtered (filter horizon).
(C) Filter horizons for a fixed period of 20 time units and different duty cycles ranging from 10\% up to 90\%. Pink shading emphasizes the location of the buffer region, and shows how the different filter horizons cover this region.
(D) Phase diagram showing the regions in which the periodic pulse train is filtered in the plane formed by $\varepsilon / \sigma$ and the input duty cycle, for a fixed period of 20 time units.
The digital and buffer regions are delimited by vertical grey dotted lines.
The insets on the left (grey) show inputs with two different duty cycles (top inset 65\%, bottom inset 20\%).
The insets on the right (colored) show the corresponding outputs for given $\varepsilon / \sigma$ ratios (with $\rho = 0.1$).
In the region between the two orange lines the filtering response depends on the specific values of the relative frequencies $\varepsilon$, $\sigma$ and $\rho$. Parameter values for all plots remain the same as in figure 3. Simulations were done with custom code in MATLAB using ODE45 function.
}
\label{fig:5}
\end{figure}

So far we have characterized the response of our timer to a single pulse of emitter activity.
However, in real-life situations such activity pulses will occur repeatedly along time, with $K_{E}$ oscillating between the $K_{E}^{\text{OFF}}$ and $K_{E}^{\text{ON}}$ values introduced in Section~\ref{sec:4} above.
In order to analyze the timer functionality in the face of repeated stimulus presentations, we consider in what follows that the emitter activation takes the form of a periodic train of pulses.
Such digital periodic signals can be characterized by their period $\tau$ and their duty cycle $d_c$, which corresponds to the fraction of the cycle during which the signal is ON (Fig.~\ref{fig:5}A).

We are interested in establishing the operation limits of the timer due to the interaction between consecutive ON pulses.
To that end, we explore in Fig.~\ref{fig:5}B the response of the system in time across all the different regions of the wire space.
Moving from left to right across the triangle, we first note that even before reaching the digital region (white region at the left) the system can already respond to the input pulses, although with a non-maximal output.
As soon as we enter into the digital region $\Phi$ (green region in Fig.~\ref{fig:5}B) the response of the system surpasses the $T_{95}$ threshold, so that the response becomes fully induced, at least during part of the input pulse.
Within that region, as we move further to the right, the turn-off lag time $t_{\text{OFF}}$ starts already to increase.
This increase becomes more evident when the system enters the buffer region $\Psi$ (magenta region in Fig.~\ref{fig:5}B), until finally the response to two consecutive pulses overlap and the signal is completely filtered out, rendering the timer unusable.

For the period and duty cycle values used in these simulations, the system starts to filter before reaching the end of the buffer region, but this will depend on the system parameters.
Figure~\ref{fig:5}C shows the location of this \textit{filter horizon} for varying values of the duty cycle and a fixed period.
The horizon gets closer to the triangle side $\overline{\rho \varepsilon}$, beyond the buffer region, for small duty cycles, and can extend into the digital region (and thus away from the region where the timer operates) for large duty cycle values.
This behavior imposes a limit in the wire space domain where the timer can function.
This domain can be visualized in Fig.~\ref{fig:5}D, which represents the regions in which the signal is either always filtered or never filtered irrespective of the specific values of the relative frequencies $\varepsilon$, $\sigma$ and $\rho$, depending on the value of the $\varepsilon / \sigma$ ratio and of the duty cycle $d_c$.
The figure shows that as the duty cycle increases, the ratio of emitters with respect to that of sink cells should decrease in order to avoid filtering the signal.
For large enough duty cycles, no part of the buffer region is free from the filtering effect, and thus the timer cannot operate.

The filtering characteristics of the system are a combined result of the duty cycle of the input signal and of the location of the multicellular system in parameter space (as defined by the ratio between the emitter and sink population densities, $\varepsilon/\sigma$). Specifically, for low enough duty cycles $d_c$ the circuit has time to respond to the square pulses without the interference of the delayed turn off typical of the buffer region. Therefore the unfiltered region in Fig.~\ref{fig:5}D is largest for low duty cycles. In turn, the larger the ratio $\varepsilon/\sigma$ the larger the buffer region, which places a stronger constraint on how small the duty cycle needs to be for filtering to be avoided.
Figure~\ref{fig:5}D thus serves as a design guide for multicellular timers in the case in which the system must respond to repetitive pulses.
In real-life situations, both the period and the duty cycle will vary in time, in which case the consortium will move across the phase diagram.
See supplementary video 3 to visualize how a culture-point lying within the digital or buffer regions processes a periodic digital signal.

\section{Discussion}
\label{sec:disc}

Here we have proposed a multicellular architecture based on three bacterial strains, an emitter, a receptor and a sink, that communicate with each other via a biological bit (bbit).
Together, these three components constitute what can be considered a chemical wire.
The sink strain degrades the bbit in the culture, allowing the system to adapt its response to the input signal in the fastest way possible.
In the absence of that strain, the communication molecule can only disappear from the culture if the media is washed or if the molecule is degraded on its own.
Controlling the bbit degradation allows fine tuning of the system dynamics, unlocking its responsiveness to a dynamical input.
To gain intuition about how the collective behavior of these three strains in culture, we have proposed a conceptual representation using trilinear coordinates to visualize the entire wire space.
This theoretical and representation framework is intended to provide a simple way to interpret the capabilities of a chemical wire implemented with our proposed architecture.
Ideally, in an experimental setting, just measuring the constants $K_E^\text{OFF}$, $K_E^\text{ON}$, $K_S$, $D_m$ and the parameters of the receptor response, $K_m$ and its Hill coefficient, would allow us to predict the range of relative frequencies for each strain in the media at which our chemical wire will work, and its performance.

Using the approach outlined above, we have identified a domain in the wire space, which we call the buffer region, that allows the wire to store information about the time during which the emitter has been in the ON state.
This represents an extra level of information, beyond the current state of the emitter, that is stored directly on the wire, operating as a timer.
This timer behavior relies on the progressive accumulation of bbit in the medium, approximately linear in time, that occurs when the circuit operates in the buffer region.
Since all cells in the consortium contribute to this accumulation, the circuit works intrinsically in a multicellular manner by averaging out the random behavior intrinsic of individual cells, leading to a more accurate temporal device \citep{Enright:1980aa,ojalvo2004}.

Our theoretical and computational results show that this biological function is robust not only to the specific values of the relative frequencies of the three strains for which the system can behave as a timer, but also in its response to repetitive stimulation pulses.
In particular, our characterization of the consortium response to a periodic train of pulses revealed that the signal is filtered out only for large enough duty cycles, and when the relative frequency of emitters is large enough in comparison with that of sink cells.
In the rest of the parameter space, the consortium performs robustly as a multicellular timer, which can be easily tuned by varying the relative frequencies of the three strains, thereby moving freely across wire space.
We note that such tuning is strongly enabled by the existence of the sink strain, which provides a well-controlled way of degrading bbit. Thus, even if that molecule is not fully stable, it is useful to have the sink strain as an element of the multicellular circuit.

Our model is based on several simplifying assumptions, including (i) that the input signal is instantaneously transduced into bbit production by the emitter strain with no intermediate steps, (ii) that bbit is directly transduced into output by the receptor strain, and (iii) that the biochemical reactions are instantaneous.
These assumptions are not limiting, since our model can be considered a starting point for models of increasing complexity and richness (although large enough delays in the production and degradation of the inducer, for instance, could substantially change the behavior of the system).
In that way, our model sets lower bounds for the wire to work properly.
Specifically, if our model predicts that a wire will not have a digital or buffer region, increasing the descriptive capability of the model by adding more complexity will not fix that issue.
On the other hand, if the model claims that either digital or buffer regions exist, a more descriptive model could be used to improve that prediction quantitatively, but the results would not change qualitatively.

We have also considered for simplicity that (1) the cells in the consortium were not proliferating and (2) the population of cells is well mixed, as in the case of a liquid culture.
The stationarity assumption 1 would not be necessary in situations where the time scales of secretion, response and degradation are much shorter than the average doubling time of the cells, as could be the case in mammalian cells, including stem cells with division times of several hours. However, in the case of a cells with high growth rate, this model looses its predictive potential. This is because the strain with the highest growth rate would dominate over the two others, displacing the point representing the culture within the triangle towards one the vertices.
Regarding assumption 2, our results could also be applicable to spatially structured populations such as adherent cells on a plate, as long as the three cell types are uniformly distributed across the plate so that any local neighborhood contains all three strains, and the bbit is a small molecule that diffuses quickly over macroscopic distances.

The analysis presented above has been phrased for clarity from a design perspective, but it is also applicable to natural multicellular circuits.
The type of cellular processes described here (production, detection and degradation of a diffusible molecule) are very generic and exist naturally in cells.
Here we show that combining these processes in the different cell types of a multicellular population could give rise to a robust and cellular timer. 
Thus, beyond their obvious implications for synthetic biology applications, our results also constitute a further instance of how cell-cell communication can potentially lead to collective functionality in cellular populations (in this case measuring time) that goes far beyond the capabilities of gene and protein circuits within individual cells.

\section*{Limitations of study}
The model presented here has some limitations that we summarize here. First, the model is aimed at describing non-growing cells. Even when, as we have already mentioned, \textit{Escherichia coli} has the ability to induce gene expression for long periods of time \citep{Gefen2014}, this capability has not been extensively exploited in synthetic biology. Thus, more research has to be done to characterize this property across model organisms. On the other hand, mammalian cells can be metabolically active without undergoing mitosis, or while going through mitosis at very low rates. This usually happens when these cells are attached to surfaces and/or to each other. Our model would apply to these types of cells, but only if the three subpopulations were homogeneously mixed. The other limitation is that our model assumes that both emitter and receptor inductions happen instantaneously. While this helps to understand the feasibility of a wire candidate, the model may require the addition of more terms if we want to capture the time dynamics in a more accurate and descriptive way.

\section*{Resource availability}
\subsubsection*{Lead contact}
Information and code requests should be directed to and will be kindly answered by the Lead Contact,
Carlos Toscano-Ochoa (biotoscano@gmail.com).
\subsubsection*{Materials availability}
This study does not comprise experimental materials.
\subsubsection*{Data and code availability}
The custom MATLAB code used in this work is available upon request to the Lead Contact.

\section*{Methods}
This is a theoretical work and hence does not provide experimental methods.

\section*{Supplemental information}
Supplemental Information consists of three videos to illustrate the model in action.

\section*{Acknowledgments}
We thank two anonymous reviewers for their insightful comments.
This work was supported by the Spanish Ministry of Science and Innovation and FEDER, under projects FIS2017-92551-EXP and PGC2018-101251-B-I00, by the ``Maria de Maeztu'' Programme for Units of Excellence in R\&D (grant CEX2018-000792-M), and by the Generalitat de Catalunya (ICREA Academia programme and grant 2017~SGR~1054).

\section*{Author contributions}
CTO: Conceptualization, Formal analysis, Software, Writing (original draft), and Visualization. JGO: Supervision, Writing (review \& editing), and Funding acquisition.

\section*{Declaration of interests}
The authors declare no competing interests.

\bibliographystyle{cas-model2-names}

\bibliography{biowire}

\end{document}